# Reflection-mode optical diffraction tomography for label-free imaging of thick biological specimens

Sungsam Kang[1], Renjie Zhou[2,*], Marten Brelen[3], Heather K. Mak[3], Peter T. C. So[1,4,5], and Zahid Yaqoob[1,*]

[1]Laser Biomedical Research Center, G. R. Harrison Spectroscopy Laboratory, Massachusetts Institute of Technology, Cambridge, Massachusetts 02139, USA
[2] Department of Biomedical Engineering, The Chinese University of Hong Kong, Hong Kong, China
[3]Department of Ophthalmology and Visual Sciences, The Chinese University of Hong Kong, Hong Kong, China
[4]Department of Mechanical Engineering, Massachusetts Institute of Technology, Cambridge, Massachusetts 02139, USA
[5]Department of Biological Engineering, Massachusetts Institute of Technology, Cambridge, Massachusetts 02139, USA
Correspondence to: rjzhou@cuhk.edu.hk & zyaqoob@mit.edu

## ABSTRACT
Optical diffraction tomography (ODT) has emerged as a powerful label-free three-dimensional (3D) bioimaging techniques for observing living cells and thin tissue layers. We report a new reflection-mode ODT (rODT) method for imaging thick biological specimens with ~ 500 nm lateral resolution and ~1 μm axial resolution. In rODT, multiple scattering background is rejected through spatio-temporal gating provided by dynamic speckle-field interferometry, while depth-resolved refractive index maps are reconstructed by developing a comprehensive inverse scattering model that also considers specimen-induced aberration. Benefiting from the high-resolution and full-field quantitative imaging capabilities of rODT, we succeeded in imaging red blood cells and quantifying their membrane fluctuations behind a turbid sample with a thickness of 2.8 scattering mean-free-paths. We further realized volumetric imaging of cornea inside an ex vivo rat eye and quantified its optical properties, including mapping the topography of Dua's and Descemet's membrane surfaces on the nanometer scale.

## INTRODUCTION
Quantitative phase imaging (QPI) has been developed to delineate structural and dynamical properties of transparent cells and thin tissues by exploring the intrinsic image contrast from refractive index (RI) and thickness variations [1]. As a label-free imaging method, QPI has enabled many unique biomedical studies [2] , such as elucidating cell growth mechanisms by quantifying mass changes at the femtogram level [3, 4], discriminating blood disease states [5-8], and probing electrical activity through measuring nanometer scale cell membrane fluctuations [9-11]. In addition, distinctive RI contrast between normal and abnormal cells and tissues have been reported for various diseases, demonstrating the potential of using RI as an intrinsic diagnostic biomarker [12-15].

Optical diffraction tomography (ODT) [16, 17] is an extension of QPI that enables volumetric imaging of biological samples by mapping their three dimensional (3D) RI maps, therefore further advancing studies in cell organelle dynamics [18, 19], pharmacology [20], immunology [21], neuroscience [22], infectious disease pathology [7], etc. In ODT, multiple 2D quantitative phase images are first measured at different illumination angles [17], wavelengths [23], sample translations [24], and so on. By solving an inverse scattering problem taking optical diffraction into its formulation [25], accurate RI reconstructions are then obtained from the 2D phase images [26, 27]. Utilizing the coherence-gating effect provided by broadband sources, white light diffraction tomography was developed for high-resolution imaging of unlabeled cells [28]. Conventional ODT methods only considered single-scattering fields by applying the first-order Born or Rytov approximation, thus limiting their applicability to studying weakly scattering objects [26]. Recently, progresses have been made to overcome this barrier by considering the higher-order scattering fields in the reconstruction models [29-32]. However, due to the limitations of the reconstruction model and apparatus, e.g., the missing-cone problem in the transmission mode ODT systems [33], ODT is still largely restricted to imaging thin objects such as cells and thin tissue slices.

The ability to image thick biological tissues in vivo is essential for many cutting-edge biological studies and clinical diagnostic applications [34]. To realize this goal in ODT, several issues must be addressed: first, a wide-field reflection-mode measurement geometry needs to be implemented; second, a comprehensive inverse scattering model that accounts for the temporal dispersion and spatial aberration of the back-scattered field from thick inhomogeneous media needs to be developed; third, the multiple scattering background needs to be suppressed to isolate the signal originated from a specific deep layer. Recently, with numerical characterization of optical diffraction and aberration, several tomographic imaging techniques have been proposed for thick tissues environments [35-38], such as rejecting the multiple-scattering fields by accurate controlling of the phase shift between the interfering waves [36].

Here, we report a new reflection-mode ODT method (rODT) that can retrieve depth-dependent RI variations and quantify structural dynamics in multiple scattering samples, while offering diffraction-limited lateral resolution and sub-micron axial resolution. rODT is based on precisely solving the inverse scattering problem of a thick sample in a speckle-correlation reflection phase microscopy (SpeCRPM) setup [39-41], where both the spatio-temporal coherence gating and the specimen-induced aberrations have been considered. Using the rODT system, we first imaged red blood cells behind a thick scattering medium with a thickness close to three scattering mean-free-paths. The high sensitivity and high-speed imaging (~ 66 frames per second (fps)) capabilities allowed us to quantify the RBC membrane fluctuations. To demonstrate its feasibility for in vivo studies, we imaged corneal structures inside an ex vivo Sprague Dawley (SD) rat eye with lateral resolution of ~500 nm and axial resolution of ~1 µm and delineated the RI values for each individual corneal layer. Surface profiles of the Dua's membrane and Descemet's membrane [42, 43], separated by 4 µm apart, were clearly resolved and their profiles were mapped with nanometer scale sensitivity.

# RESULTS

## rODT framework for thick biological samples

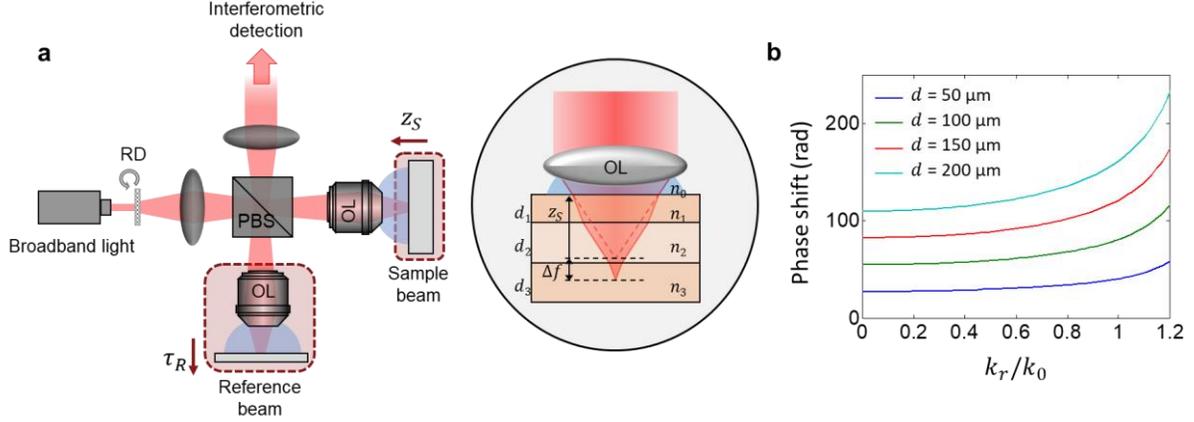

**Figure 1. Illustration of the rODT system for imaging thick samples. a**, Schematic of the rODT system. RD: rotating diffuser; PBS: polarization beam splitter; OL: objective lens. $z_S$ represents the axial position of the top surface of the sample. $\tau_R$ is the arrival time of the reference wave, which can be tuned by simultaneously moving the reference objective lens – mirror assembly. The corresponding arrows represent the direction of positive values for $z_S$ and $\tau_R$. On the right-hand side, the focus shift $\Delta f$ induced by RI mismatch inside a thick multi-layer sample is illustrated. **b**, Illustration of the resulting phase shift in a single layer as a function of normalized transverse momentum. The phase shift curves are plotted for different layer thicknesses (50 μm, 100 μm, 150 μm, and 200 μm).

As illustrated in Fig. 1a, SpeCRPM is built based on the Linnik-type interferometry, where two identical objective lenses are used in the reference and sample beam paths. To reconstruct a thick sample, we first model light propagation in the specimen by solving the wave equation under a modified first-order Born approximation and assuming the sample consists of multiple layers. Then, we derive the optical transfer function (OTF) of the rODT system to relate the sample structural information with the measurements under broadband speckle-field illumination (refer to detailed formulation of rODT framework in Supplementary Information II). When the back-apertures of both sample and reference objective lenses are uniformly filled by time-varying speckle fields, OTF can be expressed as a function of lateral spatial frequency $\mathbf{k_r} = (k_x, k_y)$, axial position of the top surface of the sample $z_S$, and arrival time of reference wave $\tau_R$ as:

$$\mathcal{T}(\mathbf{k_r}, z_S; \tau_R) = -\frac{1}{2c^2}\mathcal{F}_\omega^{-1}\left\{\omega^2|S(\omega)|^2\left[\left(\frac{P'(\mathbf{k_r},\omega)e^{iq(\mathbf{k_r},\omega)z_S}}{q(\mathbf{k_r},\omega)}\right) \star_{\mathbf{k_r}} \left(P'^*(\mathbf{k_r},\omega)e^{-iq(\mathbf{k_r},\omega)z_S}\right)\right]\right\}, \quad (1)$$

where $\mathcal{F}_\omega^{-1}[\cdot]$ stands for the inverse Fourier transform with respect to $\omega$; $\star_{\mathbf{k_r}}$ denotes for the cross-correlation with respect to the $\mathbf{k_r}$ space; $q(\mathbf{k_r},\omega) = \sqrt{n_0^2\omega^2/c^2 - k_r^2}$ is the axial projection of the scattered field wavevector inside the immersion medium with refractive index of $n_0$ with $\omega = 2\pi c/\lambda$ and $k_r = |\mathbf{k_r}| = \sqrt{k_x^2 + k_y^2}$; $|S(\omega)|^2$ is the power spectral density of the illumination source; $P'(\mathbf{k_r},\omega) = e^{i\Delta\phi(\mathbf{k_r},\omega)}P(\mathbf{k_r},\omega)$ is a complex aperture function that considers the aberration induced by the medium and $\Delta\phi(\mathbf{k_r},\omega)$ is the phase shift caused by the multi-layer sample that induces a time delay $\Delta\tau$ to the returning sample field with respect to the reference beam. At each layer interface, the refraction of the incident beam, following Snell's law, results in a focus shift $\Delta f$ along the depth as illustrated on the right-hand side of Fig. 1a. The total phase shift $\Delta\phi(\mathbf{k_r},\omega)$ from all layers is expressed as:

$$\Delta\phi(\mathbf{k_r}, \omega) = \sum_i d_i\left(\sqrt{(n_i\omega/c)^2 - |\mathbf{k_r}|^2} - \sqrt{(n_0\omega/c)^2 - |\mathbf{k_r}|^2}\right), \qquad (2)$$

where $n_i$ and $d_i$ are the RI value and thickness of $i^{\text{th}}$ layer, respectively [44]. Figure 1b shows the phase shift as a function of normalized transverse momentum $k_r/k_0$ with $k_0 = 2\pi/\lambda$, where we assumed $\lambda = 800$ nm and a single-layer structure ($n = 1.4$) whose thickness $d$ is varied to 50 μm, 100 μm, 150 μm, and 200 μm. From this graph, as expected we found that the phase shift increases with $k_r/k_0$ and $d$, especially when the value of $k_r/k_0$ is large (i.e., when a large NA objective lens is used). By taking a Fourier transform of OTF as described in Eq. (1) with respect to $\mathbf{k_r}$, the 3D point spread function (PSF) of the system $\mathcal{P}(\mathbf{r}, z_S; \tau_R)$ can be obtained.

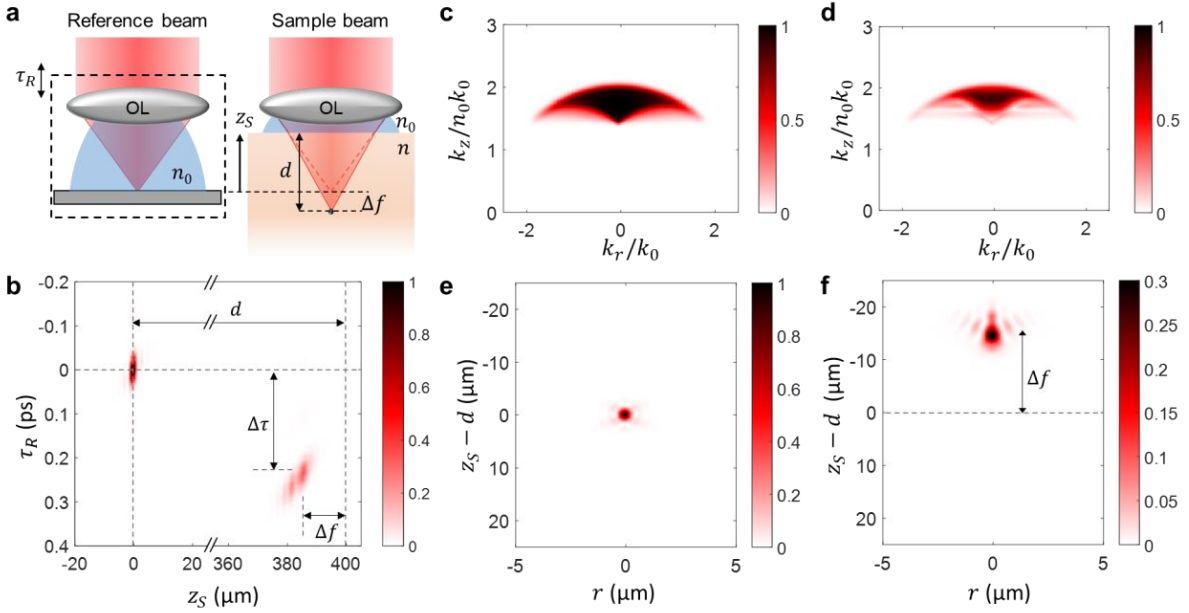

**Figure 2. Numerical simulation of PSF and OTF using the rODT model**. **a**, Illustration of the measurement scheme for the numerical simulation study. A point scatterer is positioned at depth $d$ inside a single-layer medium with an average RI value of 1.37. Due to the focus shift, the point scatterer is brought into focus by moving the sample upward to an axial location $z_S$. **b**, Point spread functions for point scatterers positioned at $d = 0$ and 400 μm. **c, d**, Optical transfer functions of the system in k-space for point scatterers positioned at $d = 0$ μm and 400 μm with $\tau_R = 0$ and 230 fs, respectively. **e, f**, Point spread functions in the real space, obtained by taking inverse Fourier transforms of the optical transfer functions shown in **c** and **d**, respectively.

To reveal the behavior of OTF and PSF in our rODT system, we performed a numerical simulation study by considering a point scatterer located at depth $d$ below the top surface of a single-layer structure whose RI value ($n$) is 1.37, as depicted in Fig. 2a. According to the actual system configuration, we assume (i) a water immersion objective lens ($n_0 = 1.33$) with NA = 1 for collecting the scattered light; and (ii) a broadband illumination source with a Gaussian-shaped power spectrum, a center wavelength at $\lambda_c = 800$ nm, and a full width at half maximum (FWHM) spectral width of $\Delta\lambda = 40$ nm. Considering the focus shift $\Delta f = z_S - d$, the point scatterer can be brought into focus by moving the sample upward to the axial location $z_S$. To optimize the interferometric signal, we need to also account for the temporal shift $\Delta\tau$ due to the optical path length delay, which can be realized by moving the reference mirror by an amount of $c\tau_R = c\Delta\tau$. It is found that the ratio of $\Delta\tau$ and the apparent PSF location $z_S$ is a function of the layer RI value ($n$), independent of the medium thickness ($d$). Therefore, we define a measurable quantity $\eta = c\tau_R/z_S = c\Delta\tau/(d + \Delta f)$ that maximize the interference signal at the depth, from which we can numerically determine the medium RI value from $\eta$. Figure 2b shows the amplitude of point spread functions ($\mathcal{P}(\mathbf{r} = 0, z_S, \tau_R)$) for two-point scatterers at $d = 0$ μm and 400 μm, respectively. For the point scatterer at the surface ($d = 0$ μm), imaging is optimized by setting $z_S = 0$ μm and $\tau_R = 0$ ps. For the point scatterer at $d = 400$ μm, due to the RI difference of the medium and the sample, there is an apparent focus shift with $\Delta f = -14.1$ μm and a considerable temporal shift with $\Delta\tau = 0.23$ ps. The scatterer is brought into focus by re-positioning the sample at $z_S = 386$ μm, while the interferometric signal is optimized by delaying the reference mirror with $\tau_R = 0.23$ ps. A more detailed discussion, including the focus shift, temporal shift, and RI reconstruction, can be found in Methods section and Section III and IV of the Supplementary Information.

Figure 2c shows the amplitude of the cross-section of the OTF $\mathcal{T}(k_r, k_z; \tau_R = 0)$ rescaled in ($k_r/k_0, k_z/n_0k_0$) space for the point scatterer is at the surface ($d = 0$ μm). As expected, OTF has a high spatial frequency distribution along $k_z$ because of the reflection-mode geometry. It is also noted that the OTF shows a "soft" boundary because of broadband illumination, where the width of the soft region is determined by the spectral bandwidth of the light source. Note that for the monochromatic case, OTF exhibits a uniform distribution within a discrete boundary determined by the NA of the system [41]. Furthermore, due to the use of broadband illumination, the effective bandwidth along $k_z$ axis is increased, thus resulting in an increase in axial resolution. By taking an inverse Fourier transform, we can obtain the PSF $\mathcal{P}(r, z_S; \tau_R = 0)$ which is presented in Fig. 2e. Then the spatial resolution of the system can be estimated from the FWHM values of the intensity PSF $|\mathcal{P}(r, z_S; \tau_R = 0)|^2$ along the lateral and axial directions, which are 300 nm and 1.2 μm, respectively. $\mathcal{T}(k_r, k_z; \tau_R)$ is also computed for the point scatterer at $d = 400$ μm with $\tau_R = \Delta\tau = 0.23$ ps, as shown in Fig. 2d. The shape of OTF is distorted due to the spherical aberration induced by the sample, effectively reducing the spatial frequency coverage range. As a result, the corresponding PSF (Fig. 2f) is also broadened and asymmetrically distorted with a negative focus shift $\Delta f = -14.1$ μm. For this case, the spatial resolution along the lateral and axial dimensions are estimated to be 400 nm and 2 μm, respectively.

**Quantitative volumetric imaging of tissue phantom**

To validate our rODT model and demonstrate the quantitative phase imaging capability within a thick scattering medium, we first assembled a sample (Fig. 3a) that consists of three distinct layers: tissue phantom layer comprising 2% intralipid in gelatin (L1), glass coverslip layer (L2), and a buffer solution layer where human red blood cells (RBCs) are suspended (a thick glass substrate is placed at the bottom). Note that we placed another coverslip on top of the scattering tissue phantom to prevent water swelling at the gelatin layer due to the use of a water-immersion objective lens, while a similar coverslip was introduced in the reference arm to cancel the spherical aberration of the top coverslip on the PSF. With the SpeCRPM system, we first found the PSF locations in the ($z_S, \tau_R$) space for each of the four interfaces associated with layers L1, L2 and L3, as shown in Fig. 3b. Next, we retrieved RI value ($n$) and thickness ($d$) of each layer by using the look-up table $L(z_S, \tau_R)$ as shown in Fig 5, as illustrated in Figs. 3c-d. We found $n = 1.512 \pm 0.006$ and $d = 149.8 \pm 1.2$ μm for the middle coverslip layer (L2), which is in a good agreement with the manufacturer's specifications. We also identified the RI value and thickness of gelatin layer L1 ($n = 1.346 \pm 0.002$, $d = 452.3 \pm 1.3$ μm) and that of the buffer solution ($n = 1.335 \pm 0.003$, $d = 124.7 \pm 1.6$ μm) in layer L3. Next, we acquired the volumetric image of the sample by scanning both the sample and reference arms, while the optimum arrival time ($\tau_R$) of the reference arm at every depth ($z_S$) is obtained according to the solid line in Fig. 3b. The cross-sectional image at $y = 0$ plane is shown in Fig. 3e (left), where all interfaces of the three layers are clearly visible. The signal decay curve from the intralipid scatterers in the first layer

(L1) is also shown in Fig. 3e (right). From the exponential decay of the signal, we characterized the scattering mean-free-path $l_s$ of layer L1 to be ~160 µm. The quantitative phase image of individual RBCs at the bottom interface of the third layer (L3) at 700 µm in depth is presented in Fig. 3f. Despite the presence of scattering tissue phantom with a thickness of ~$2.8l_s$, the RBCs are clearly seen in the presented phase map. We also recovered the height-map of an RBC, indicated by dotted square box 1 in Fig. 3f, by assuming an RI contrast of 0.06 [7]. In addition, we measured the time-lapse the RBC membrane dynamics at 66 fps. The root-mean-square (RMS) displacements of three different RBCs (indicated by dotted square box 1, 2, and 3 in Fig. 3f) were analyzed (Fig. 3h). The observed mean RMS displacement of RBCs is around 100 nm which is significantly larger than that of the background at around 40 nm, thus demonstrating the potential of our system for quantifying cellular rheology in a highly scattering sample by means of quantitative phase imaging [5-8]. It is noteworthy that the RMS displacement of background is higher than our previous result [39], which is mainly due to the remaining multiple scattering noise that passes the spatio-temporal coherence gating of our system.

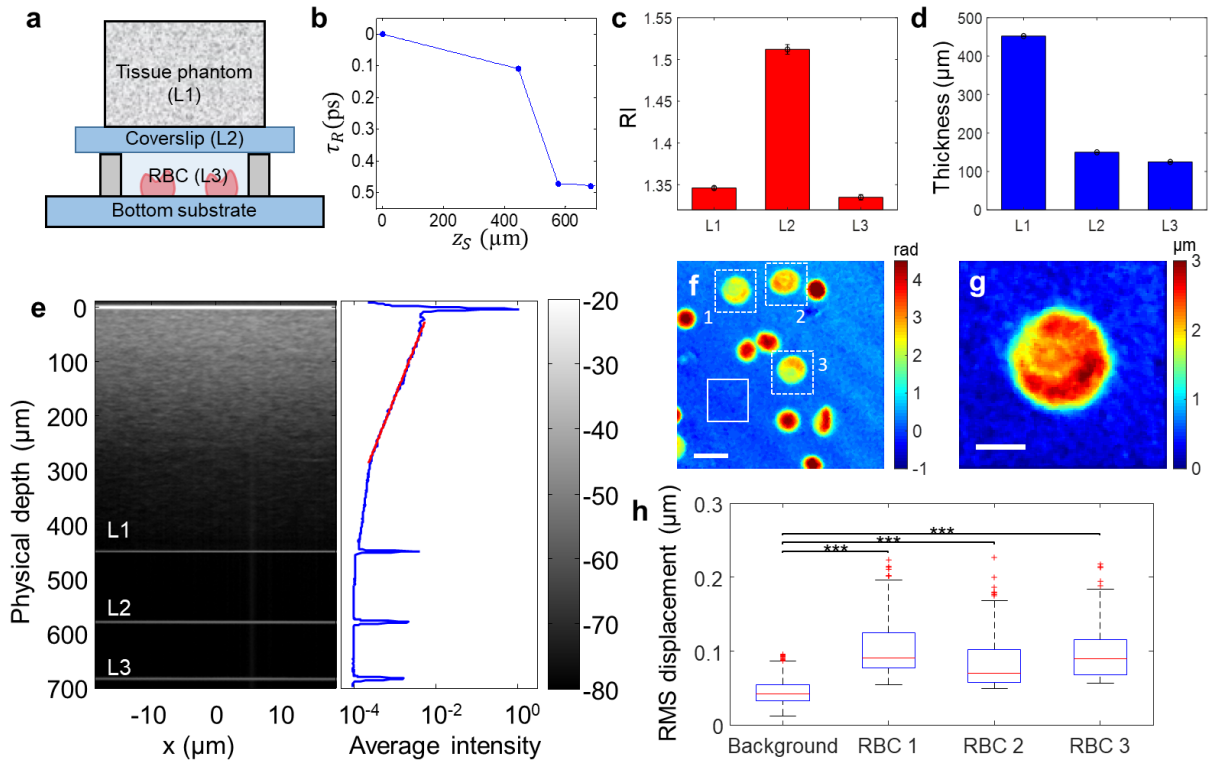

**Figure 3. Quantitative phase imaging of RBCs at the bottom of a scattering tissue phantom. a**, Schematic diagram of the assembled sample composed of three layers: scattering tissue phantom layer (L1), coverslip layer (L2), and buffer solution layer (L3) where RBCs are suspended. **b**, PSF locations in the $(z_S, \tau_R)$ space for the four interfaces associated with layers L1, L2 and L3 of the sample. **c-d**, Retrieved RI value (*n*) and thickness (*d*) of the three layers. **e**, Normalized intensity cross-sectional image (left) and average intensity cross-section curve of the sample (right) with reference path length correction according to the $(z_S, \tau_R)$ relation shown in **b**. **f**, Quantitative phase image of RBCs at the bottom interface of layer L3. Scale bar: 10 µm. **g**, Height-map of RBC 1 as indicated in **f**. Scale bar: 4 µm. **h**, RMS displacement of the three RBCs labeled in **f** (dashed white box) and the background (solid white box). The red line within each box represents the median, while the lower and upper boundaries of the box indicate the first and third quartiles. Whiskers represent 1.5 interquartile range. Red markers represent outliers beyond 1.5 interquartile range. Two-sample t-test suggests that there are statistically significant differences (***$p<0.001$) between the RMS-displacements of the three RBCs and the background.

Note that when dealing with real tissue samples, discrete interfaces may not be available. Instead, individual cells and intracellular structures can be used as interfaces, where back-scattered fields undergo angle-dependent phase shifts in their way in and out by the average RI of the upper tissue layer. In such a scenario, the signal is distributed in the $(z_S, \tau_R)$ space around a linear line whose slope $\eta/c$ is determined by the average RI value ($n$) at the corresponding depth. More specifically, we will first determine $\tau_R$ at depth $z_S$ by moving the reference mirror until an optimum interference signal is reached. Then, we will retrieve the average RI value ($n$) and thickness ($d$) of the volume between the top surface and the focal plane using the look-up table. By successively moving the sample along the axial direction and determining $\tau_R$ at each depth $z_S$, we can eventually obtain the RI value for each depth for a real tissue sample.

**Volumetric imaging of rat corneal tissue**

To demonstrate the potential of rODT for in vivo imaging, we performed ex vivo imaging of the corneal tissue on an intact SD rat eye (refer to Methods section for the sample preparation). We first obtained the volumetric image by scanning the sample along the depth, from which we virtually divided the volume into six regions along the depth dimension. Then, the optimal $\tau_R$ at each interface was found according to the PSF location in the $(z_S, \tau_R)$ space, which is further used to retrieve the RI value at each layer (left panel of of Fig. 4a) based on the look-up table (refer to Supplementary Information section VI for more details). After that, we rescanned the corneal tissue by correcting the temporal shift induced by the sample according to the PSF locations in the $(z_S, \tau_R)$ space (the process is similar to obtaining Fig. 3b-c). Figure 4a shows the corresponding cross-sectional image after scanning both the sample as well as the reference path length. As shown in Fig. 4a, the corneal tissue clearly shows three distinct layers with apparently different RI values: epithelium layer with average $n = 1.37$ and $d = 32$ µm, stroma layer layer with average $n = 1.35$ and $d = 138$ µm, and endothelium layer with average $n = 1.39$ and thickness $d = 8$ µm, respectively. The measured RI and thickness values for these layers in general agree well with the previously report [45]. In addition, we identified various detailed features of the corneal tissue, such as the interior stroma layer at depth 133 µm where nerve fiber structures are shown (Fig. 4i). Specifically, as shown in Fig. 4b, we found two distinct layers ($L_1$, $L_2$) that correspond to the Dua's membrane [42] and the Descemet's membrane [43], located at $d = 164$ µm and 170 µm, respectively. The enface amplitude and quantitative phase images of each layer are shown in Figs. 4c-d and Figs. 4f-g, respectively. The amplitude maps of $L_1$ and $L_2$ show relatively uniform reflection by the smooth membrane surface, while the phase maps reveal apparent morphological variations of the Dua's and Descemet's membranes. From the phase maps, we obtained the height maps of $L_1$ and $L_2$ (Fig. 4e-h), which show variations of 103 nm and 241 nm, respectively. Note that the height is converted from the phase using the calculated RI value (~1.341) for the lower stroma layer as illustrated in Fig. 4a.

To further improve the volumetric image contrast and resolution, we performed a 3D deconvolution based on the Richardson-Lucy algorithm [46, 47]. For this purpose, we first virtually divided the whole 3D volume into multiple 20 µm-thick sub-volumes along the depth of the tissue sample and computed the PSF for each sub-volume according to the measured RI distribution. Next, we performed 3D deconvolution on each sub-volume with the corresponding PSF. Finally, all the deconvolved sub-volumes were stitched together to render a whole new 3D volume with optimized image contrast and spatial resolution. Comparing enface images of the interior stroma layer before and after the deconvolution (Figs. 4i&j), there is a clear improvement in the image contrast and lateral resolution for the submicron features. In Fig. 4k, we plot cross-sections across a nerve fiber, indicated by dashed lines in Figs. 4i-j, before and after the deconvolution. The FWHM of the nerve fiber was measured to be 584 nm as opposed to 751 nm before the deconvolution, showing ~30% improvement in resolution.

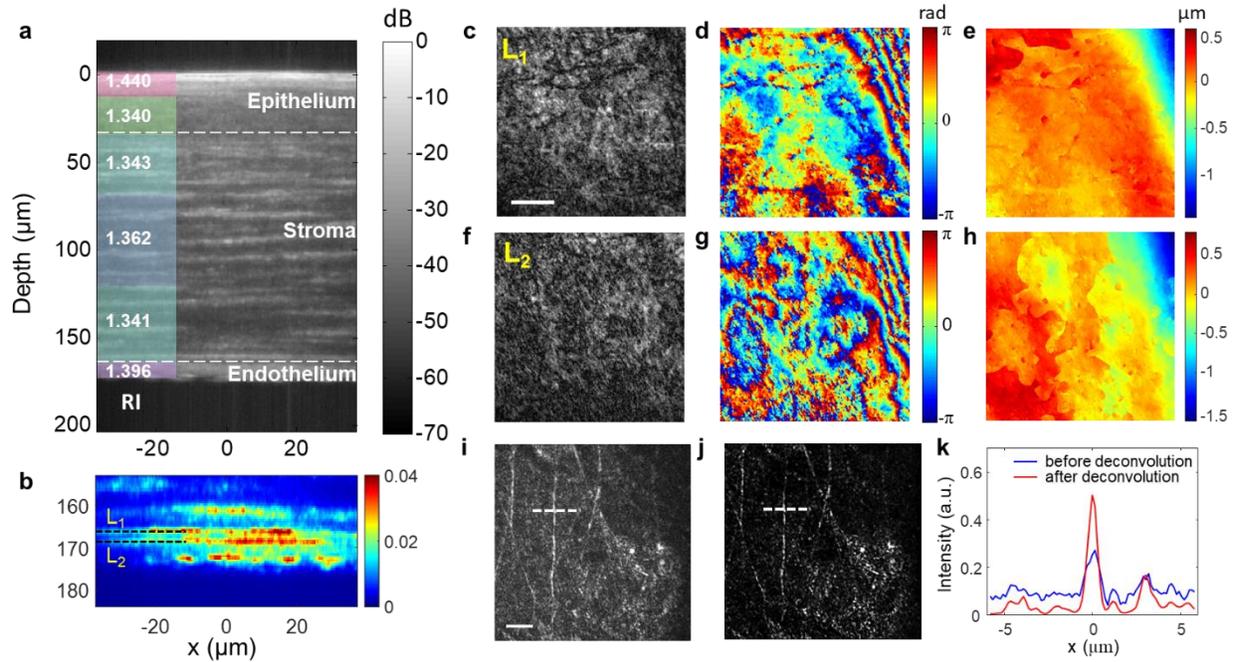

**Figure 4. Volumetric imaging of a fixed SD rat corneal tissue**. **a**, Sectional intensity image of a rat corneal tissue with log scale color mapping (unit in dB). Average RI value of each layer is displayed on the left side. **b**, Magnified image of **a** at depth 155 to 185 μm with linear scale color mapping. Two distinct layers associated with the Dua's membrane and the Descemet's membrane are clearly identified and labeled as $L_1$, and $L_2$, respectively. **c-h**, Amplitude, phase, and height maps of layers $L_1$ and $L_2$. Scale bar: 10 μm. **i-j**, Amplitude images at depth $d = 133$ μm before and after 3D deconvolution, respectively. Scale bar: 10 μm. **k**, Cross-section along the dotted white line in **i-j**. Blue line: before 3D deconvolution; Red line: after 3D deconvolution.

## Summary and outlook

Based on broadband illumination and dynamic speckle-field interferometric microscopy, rODT has been developed to realize epi-mode volumetric imaging of bulk biological tissue samples through precise mapping of the 3D amplitude and phase distributions of the sample back-scattered fields. Various optical and biophysical parameters, including the depth-resolved RI and 3D PSFs inside the sample, are retrieved with the proposed rODT. The performance of rODT was first verified by mapping the RI values and layer thicknesses of a scattering tissue phantom and imaging live RBCs placed underneath the tissue phantom. Furthermore, we demonstrated high-resolution 3D imaging of an ex vivo SD rat corneal tissue by identifying the RI values and thicknesses of its distinct layers. High depth-selectivity of the system revealed detailed surface profiles of the Dua's and Descemet's membranes in the cornea which were separated 4 μm apart. From these discrete reflective surfaces, depth profiling at the nanometer scale is possible. By developing a depth-resolved 3D deconvolution algorithm, we improved the spatial resolution by ~30%.

In the current SpeCRPM setup, we can fully account for temporal delay in the specimen, thus improving the interferometric signal contrast and image stack signal-to-noise ratio. Although we can quantity PSF distortion due to spherical aberration, the spatial resolution of the image stack will unavoidably degrade at deeper layers. The numerical deconvolution approach may partially recover the lost information, but in a future system design one may need to include an adaptive optics module, which can fully account for specimen induced aberration so as to better maintain the spatial resolution and signal-to-noise level, even at deeper tissue layers. While the current rODT model can provide 1D depth-resolved RI as well as high-resolution volumetric imaging of the thick biological tissue samples, a full recovery of the 3D RI distribution is not straight forward. There is a possibility to fully recover the 3D RI distribution by using the knowledge of 1D RI provided by our model in combination with the regularization methods, such as Gerchberg-Papoulis spectral extrapolation technique [48]. We envision that the RI maps of thick biological tissues can be potentially used to derive intrinsic biomarkers for disease diagnosis in a label-free fashion [14, 15]. With the high-resolution and full-field imaging capabilities, rODT may potentially offer more accurate disease diagnosis and biological investigations in vivo.

## METHODS

**Experimental setup for acquiring volumetric data**
The basic working principle of SpeCRPM (Fig. 1a) can be briefly described as follows. Linearly polarized broadband dynamic speckle-field, generated by a rotating diffuser (RD), is divided into sample and reference beams using a polarization beam splitter (PBS). A quarter wave plate is placed in each arm at 45º, allowing for the cross-polarized back-scattered fields from the sample and reference arms to be combined collinearly at the same PBS and guided into the detection arm. Furthermore, off-axis holography is implemented for single-shot and wide-field imaging [39, 40]. The interferogram is recorded using a CMOS camera (Flea3, Point Gray) placed in the image plane conjugate with the sample. Two motorized translational stages are used in the sample and the reference arms to control the axial location $z_S$ of the sample and arrival time $\tau_R$ of the reference beam. Details of SpeCRPM is further discussed in Sec. I of the Supplementary Information.

**Look-up table for RI and thickness reconstruction**

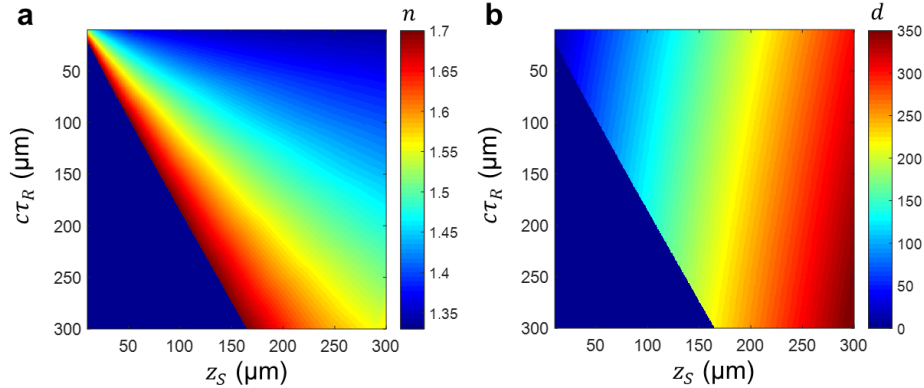

**Figure 5. Look-up tables for RI and thickness reconstruction.** **a-b**, Look-up tables $L(z_S, \tau_R)$ for retrieving $n$ and $d$ values of layered structures. Blank region in both figures corresponds to $n > 1.7$. We assume NA = 1, center wavelength $\lambda_c = 800$ nm, and spectral bandwidth $\Delta\lambda = 40$ nm.

Figure 2 presents a numerical simulation of the rODT model for a single-layer sample with $n = 1.37$ and $d = (0, 400)$ µm. To generalize to an arbitrary sample, we first consider a multi-variable function of shifted PSF locations as a function of the RI value ($n$) and the thickness ($d$) as $[z_S, \tau_R] = h(n, d)$. Next, we numerically determine the functional form of $h(n, d)$ using Eq. (1), with specific system parameters such as the power spectral density of the illumination source, NA of the system, and RI of the immersion medium. Then, we calculate a look-up table of the RI value and the thickness of the sample as a function of shifted PSF position within the sample as $[n, d] = L(z_S, \tau_R)$, where $L$ represent an inverse mapping of $h$ as shown in Fig. 5. With the look-up table, we can obtain the RI value and thickness of an arbitrary medium by measuring the optimum reference arrival time which maximizes the signal at a particular depth. Details of the look-up table is further discussed in Sec. III – IV in the Supplementary Information.

## Sample preparation

**Experimental animals:** All experimental procedures were approved by The Chinese University of Hong Kong Animal Experimentation Ethics Committee and Hong Kong Department of Health, which adhere to The International Guiding Principles for Biomedical Research Involving Animals and The Hong Kong Code of Practice for Care and Use of Animals for Experimental Purposes. Sprague-Dawley (SD) rats were fed standard diet ad libitum and housed in a 12-hr light/12-hr dark light cycle.

**Rat corneal tissue sample:** At 30-days-old, Sprague Dawley (SD) rats were anaesthetized with 100 mg/kg ketamine and 9 mg/kg xylazine and perfused using 4% paraformaldehyde (PFA). The eyeballs were harvested and immersed in 4% PFA at 4°C overnight for chemical fixation. On the next day, eyeballs were washed using phosphate buffered saline (PBS) three times before dissection of corneal tissue.

**Acknowledgment**
S.K., P.T.C.S., and Z.Y. acknowledge support from National Institutes of Health (NIH) funding 5-P41-EB015871-27, and Hamamatsu Corporation. P.T.C.S. further acknowledges support from the Singapore–Massachusetts Institute of Technology Alliance for Research and Technology (SMART) Center, Critical Analytics for Manufacturing Personalized-Medicine IRG. P.T.C.S. and Z.Y. further acknowledge support from NIH R01DA045549, R21GM140613-02, R01HL158102. R.Z. acknowledges support from Croucher Innovation Awards 2019 (Grant No. CM/CT/CF/CIA/0688/19ay), Hong Kong Innovation and Technology Commission (ITS/148/20 and ITS/178/20FP); and The Chinese University of Hong Kong Research Sustainability of Major RGC Funding Schemes – Strategic Areas.

**Author contributions**
P.T.C.S and Z.Y. conceived the project. R.Z., P.T.C.S, Z.Y., and S.K. proposed the idea and design of the system. S.K. and R.Z. developed the theoretical framework. S.K. built the setup, performed experiments, and prepared the figures. All authors analyzed and interpreted the data. M.B. and H.M. provided the rat eye sample. S.K., R.Z., P.T.C.S, and Z.Y. prepared the manuscript. All authors reviewed and edited the manuscript.